\documentclass[aps,prl,reprint,amsmath,amssymb,longbibliography]{revtex4-2}

\usepackage[utf8]{inputenc}
\usepackage[T1]{fontenc}
\usepackage{amsthm}
\usepackage{amsmath}
\usepackage{amssymb}
\usepackage{braket}
\usepackage{xcolor}
\usepackage[colorlinks=true, citecolor=blue, linkcolor=blue, urlcolor=blue]{hyperref}
\usepackage{tikz}
\usetikzlibrary{calc, arrows.meta, intersections}

\newtheorem{theorem}{Theorem}
\newtheorem{lemma}{Lemma}

\newcommand{\tr}{\text{Tr}}

\begin{document}
\title{Resourcefulness without Resource: Geometric Origins and Robustness}

\author{Jingsong Ao}
\author{Aby Philip}
\author{Alexander Streltsov}
\email{streltsov.physics@gmail.com}
\affiliation{Institute of Fundamental Technological Research, Polish Academy of Sciences, Pawi\'{n}skiego 5B, 02-106 Warsaw, Poland}
\begin{abstract}
A prevailing intuition holds that quantum protocols using only free states confer no operational advantage. This intuition is contradicted by free-state discrimination gaps in which restricted measurements fail to optimally distinguish even orthogonal free states. Known instances include nonlocality without entanglement and, more recently, nonstabilizerness without magic. We trace these examples to a single convex-geometric mechanism: whenever the set of free measurements is closed, convex, and strict subset the set of all measurements, and the free states is a convex set with an interior, a gap-witnessing ensemble can be drawn entirely from the free states. The resulting gap is operationally rigid: no finite-dimensional assistance---catalyst or quantum memory---can asymptotically improve the discrimination rate beyond the single-shot restricted limit. By contrast, non-free ensembles admit memory-assisted attacks that fully erase the gap, exposing a sharp operational asymmetry between free and resource-carrying ensembles.
\end{abstract}
\maketitle
\paragraph{Introduction}
A central objective of modern quantum science is to identify and quantify the physical properties that unlock quantum advantages. Quantum resource theories (QRTs)~\cite{chitambar2019resource} provide a unifying framework for this pursuit, treating properties such as entanglement, coherence, or magic as a resource to be used sparingly. States not containing any resources are called free states, and operations that generate no resources are called free operations. Naturally from this framing, we expect that operations and protocols that consume no resources, i.e., use only free states and operations, should confer no operational advantage over using only classical resources.

This intuition was overturned by Bennett et al.'s discovery of ``nonlocality without entanglement'' (NLWE)~\cite{bennett1999nlwe}: there exists a set of orthogonal separable states (free states) that cannot be perfectly distinguished using local operations and classical communication (LOCC),  or the free measurements in the resource theory of entanglement. Since this initial discovery, examples of such ensembles have been reported in the bipartite~\cite{Berry_Groisman_2001,Yu_2012,yang2015characterizing} and multipartite entanglement~\cite{divincenzo2003unextendible,4957660,PhysRevA.98.022303,PhysRevA.93.032341,PhysRevA.95.052344}. An important application of such states is that a bit of classical information can be encoded into orthogonal separable bipartite or multipartite states that can be prepared using LOCC, yet are almost indistinguishable by any LOCC procedure. This phenomenon is often referred to as quantum data hiding~\cite{PhysRevLett.86.5807,PhysRevLett.89.097905,DiVincenzo_2002,Hayden_2004,Aubrun_2015} and can be thought of as an information-theoretic primitive for secret sharing. 

More recently, Kwon~\cite{kwon2025} established an analogous phenomenon in the resource theory of magic called  ``nonstabilizerness without magic'': there exist orthogonal stabilizer states that cannot be perfectly distinguished by stabilizer operations. Both phenomena reveal the possibility of free state ensembles in certain resource theories exhibiting a \emph{free-state discrimination gap}: A set of free states that are strictly less distinguishable when using only free measurements than if all possible measurements were allowed.

The presence of such similar phenomena in two structurally distinct resource theories warrants further investigation. Prior results~\cite{bae2015discrimination,Oszmaniec2019operational,takagi2019general} have shown that there exist ensembles of states that are strictly less distinguishable using free measurements than if all measurements were available. But, in these works, the resource content of such ensembles was not explored. We can then ask: Is there a unifying explanation for free-state discrimination gaps that may occur in different resource theories, possibly from the structure of resource theories in general? The possibility of free states that cannot be distinguished using free measurements could give rise to a generalization of quantum data hiding~\cite{PhysRevLett.86.5807,PhysRevLett.89.097905,DiVincenzo_2002,Hayden_2004,Aubrun_2015} .  

Another question we consider is, whether the free-state discrimination gap can be overcome within sequential protocols using quantum catalysis~\cite{jonathan1999entanglement,datta2023catalysis,kondra2021catalytic} or quantum memory~\cite{philip2025robustnessquantumdatahiding}. Both quantum catalysts and quantum memory are auxiliary finite-dimensional quantum systems. Previously in~\cite{philip2025robustnessquantumdatahiding}, the authors explored this question for the resource theory of entanglement. It has been shown that neither quantum catalysts nor quantum memory can increase the optimal discrimination probability in the case of quantum data hiding.

In this work, we identify a sufficient condition on resource theories for the existence of \emph{free state ensembles} that exhibit the free-state discrimination gap.
This condition unifies NLWE and nonstabilizerness without magic and predicts analogous instances across resource theories satisfying the same hypotheses.

We further prove that no finite-dimensional assistance, quantum catalysis nor memory, can asymptotically close the free-state discrimination gap. However, in the case of non-free ensembles, finite-dimensional assistance in the form of quantum memory can overcome the discrimination gap between free and general measurements.

\paragraph{Preliminaries}

Let $\mathcal{H}$ denote a finite-dimensional Hilbert space, and let $\mathrm{Herm}(\mathcal{H})$ and $\mathcal{D}(\mathcal{H})$ be the sets of Hermitian operators and density matrices (states) on $\mathcal{H}$, respectively. A quantum resource theory is characterized by a set of free states $\mathcal{F} \subseteq \mathcal{D}(\mathcal{H})$ and a set of free operations (channels) $\mathcal{O}$ that are non-resource-generating, meaning $\Lambda(\sigma) \in \mathcal{F}$ for all $\sigma \in \mathcal{F}$ and $\Lambda \in \mathcal{O}$~\cite{chitambar2019resource}. 

A quantum state discrimination task is defined by an ensemble $\mathcal{S} = \{p_i, \sigma_i\}_{i=1}^N$, where the states $\sigma_i$ are prepared with prior probabilities $p_i$ such that $\sum_{i=1}^N p_i = 1$. To distinguish these states, one performs a measurement described by a positive operator-valued measure (POVM), i.e.\ a tuple of operators $\mathcal{M} = \{M_i\}_{i=1}^N$ satisfying $M_i \ge 0$ and $\sum_{i=1}^N M_i = \mathbb{I}$~\cite{qsd}. For a binary ensemble $\{1/2, \rho; 1/2, \sigma\}$, Helstrom's theorem~\cite{helstrom1969quantum} states that the optimal success probability is exactly $1/2 + \|\rho - \sigma\|_1 / 4$, where $\|\rho - \sigma\|_1 = \tr|\rho - \sigma|$ is the trace distance.  

The average success probability of distinguishing the ensemble $\mathcal{S}$ using a POVM $\mathcal{M}$, $P(\mathcal{S}, \mathcal{M})$, is as follows:
\begin{equation}
    P(\mathcal{S}, \mathcal{M}) = \sum_{i=1}^N p_i \tr[\sigma_i M_i].
\end{equation}
We denote the set of all valid $N$-outcome POVMs on $\mathcal{H}$ by $\mathrm{POVM}_N(\mathcal{H})$, and $\mathbb{M} \subset \mathrm{POVM}_N(\mathcal{H})$ denotes the restricted set of free measurements allowed by the resource theory, for example, LOCC measurements in entanglement theory, or stabilizer measurements in the resource theory of magic. The maximum success probability under restricted measurements, $\mathbb{M}$, is as follows:
\begin{equation}
    P_{\mathbb{M}}(\mathcal{S}) = \sup_{\mathcal{M} \in \mathbb{M}} P(\mathcal{S}, \mathcal{M}).
\end{equation}
The maximal success probability under arbitrary measurements is $P_{\mathrm{All}}(\mathcal{S}) = \sup_{\mathcal{M} \in \mathrm{POVM}_N(\mathcal{H})} P(\mathcal{S}, \mathcal{M})$. A state discrimination gap exists if 
\begin{equation}
    P_{\mathbb{M}}(\mathcal{S}) < P_{\mathrm{All}}(\mathcal{S}).
\end{equation}
We call it a \emph{free-state discrimination gap} when the ensemble $\mathcal{S}$ that exhibits this discrimination gap consists entirely of free states.

\paragraph{Geometric Origin of the Gap}

As mentioned prior, a free-state discrimination gap exists in both the resource theory of entanglement~\cite{bennett1999nlwe,PhysRevLett.86.5807,PhysRevLett.89.097905,DiVincenzo_2002,Hayden_2004,Aubrun_2015} and of magic~\cite{kwon2025}. Since these two resource theories are quite different, we aim to provide a unifying criterion that explains both phenomena and will help identify other resource theories that exhibit such a free-state discrimination gap. The following theorem provides a simple geometric criterion that is sufficient for the existence of a free-state discrimination gap. For simplicity, we shall focus on resource theories where the set of free measurements $\mathbb{M}$ is closed and convex.

\begin{theorem}\label{thm1}
If $\mathcal{F}$ has an interior, then for any resourceful POVM $\mathcal{M}^*\notin\mathbb{M}$ there exists a state ensemble $\mathcal{S}=\{p_i,\sigma_i\}$ with every $\sigma_i$ a free state such that
\begin{equation}\label{eqn:theorem_1}
    P(\mathcal{S},\mathcal{M}^*) > P_{\mathbb{M}} (\mathcal{S}).
\end{equation}
\end{theorem}

For a detailed proof of Theorem~\ref{thm1}, see Supplementary Material. We provide a brief outline of proof here. To begin, note that, the set of free measurements $\mathbb{M}$ is a closed convex proper subset of the set of all possible POVMs, $\mathrm{POVM}_N(\mathcal{H})$, and hence any resourceful POVM $\mathcal{M}^*\notin\mathbb{M}$ can be separated from $\mathbb{M}$ by a hyperplane using the strict separation theorem~\cite{boyd2004convex}. A similar argument was used by Oszmaniec and Biswas in~\cite{Oszmaniec2019operational}. The hyperplane separating  $\mathcal{M}^*\notin\mathbb{M}$ and $\mathbb{M}$ is characterized by a tuple of Hermitian operators $\mathcal{A}=(A_1,\dots,A_N)$. Using the Hermitian operators in $\mathcal{A}$ and the fact that $\mathcal{F}$ has an interior, we can construct an ensemble \emph{entirely from free states}. 
To do this, we begin by shifting the normal operators $A_i$ by a large multiple of $\gamma$, a suitable point in the interior of $\mathcal{F}$, to get  $B_i=A_i+t\gamma$. We then normalize each $B_i$, yielding an ensemble of free states that satisfies \eqref{eqn:theorem_1}.

The sufficient condition from Theorem~\ref{thm1}, that the set of free states $\mathcal{F}$ has an interior, is satisfied by several resource theories: separable states~\cite{werner1989quantum}, PPT states~\cite{Peres_1996}, the stabilizer polytope in the resource theory of magic~\cite{Veitch_2014}, and unsteerable~\cite{Gallego_2015} or local-hidden-variable states, all form convex bodies with an interior. Consequently, Theorem~\ref{thm1} furnishes a single convex-geometric mechanism that \emph{subsumes} both nonlocality without entanglement~\cite{bennett1999nlwe} and nonstabilizerness without magic~\cite{kwon2025}: each is an instance of the free-state ensemble it guarantees, recovered here by an argument considerably simpler than the original explicit constructions. It also strengthens earlier existence results. Prior results~\cite{Oszmaniec2019operational,takagi2019general} showed that any strict measurement restriction induces a gap for \emph{some} ensemble, the witnessing states there need not be free. By explicitly leveraging the fact that the free states possess an interior, Theorem~\ref{thm1} guarantees that this gap can be witnessed \textit{exclusively} by  free states. 

The existence of free-state discrimination gap in a broad range of resource theories invites us to analyze the existence of generalized quantum data hiding schemes where classical bits are encoded in an ensemble of free quantum states such that resource constrained users cannot distinguish these states. Similar to quantum data hiding, generalized quantum data hiding can be thought of as a cryptographic primitive.

\paragraph{Robustness of Free-State Ensemble Discrimination Against Finite-Dimensional Assistance}

Since generalized quantum data hiding is realized on free states that cannot be distinguished using free measurements, a question that arises is whether access to small amount of resources can help distinguish such free states and thus ``unhide'' the encoded data. In~\cite{philip2025robustnessquantumdatahiding}, it has been shown in the case of resource theory of entanglement that neither entanglement available via quantum catalysts nor quantum memories can increase the probability of distinguishing separable states using LOCC alone. We now ask whether such a result holds true in other resource theories. 

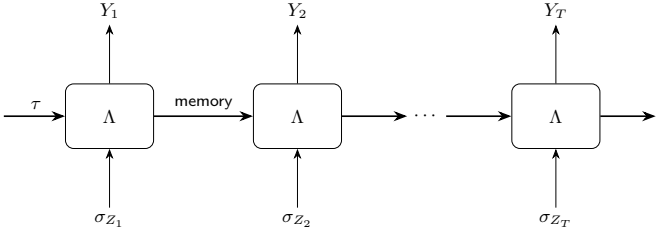
\begin{figure}[t]
    \centering
    \resizebox{\columnwidth}{!}{%
    \begin{tikzpicture}[>=Stealth, font=\sffamily]
        \node[draw, rounded corners, minimum width=1.5cm, minimum height=1.1cm] (A) at (0,0) {$\Lambda$};
        \node[draw, rounded corners, minimum width=1.5cm, minimum height=1.1cm] (B) at (3.2,0) {$\Lambda$};
        \node (D) at (5.4,0) {$\cdots$};
        \node[draw, rounded corners, minimum width=1.5cm, minimum height=1.1cm] (C) at (7.6,0) {$\Lambda$};
       
        \draw[->, thick] (-1.8,0) -- (A.west);
        \node[above] at (-1.25,0) {$\tau$};
        \draw[->, thick] (A.east) -- (B.west);
        \draw[->, thick] (B.east) -- (D.west);
        \draw[->, thick] (D.east) -- (C.west);
        \draw[->, thick] (C.east) -- (9.3,0);
        \node[above] at (1.6,0) {\footnotesize memory};

        \draw[->] ([yshift=-1cm]A.south) -- (A.south);
        \node[below] at ([yshift=-1cm]A.south) {$\sigma_{Z_1}$};
        \draw[->] ([yshift=-1cm]B.south) -- (B.south);
        \node[below] at ([yshift=-1cm]B.south) {$\sigma_{Z_2}$};
        \draw[->] ([yshift=-1cm]C.south) -- (C.south);
        \node[below] at ([yshift=-1cm]C.south) {$\sigma_{Z_T}$};

        \draw[->] (A.north) -- ([yshift=1cm]A.north);
        \node[above] at ([yshift=1cm]A.north) {$Y_1$};
        \draw[->] (B.north) -- ([yshift=1cm]B.north);
        \node[above] at ([yshift=1cm]B.north) {$Y_2$};
        \draw[->] (C.north) -- ([yshift=1cm]C.north);
        \node[above] at ([yshift=1cm]C.north) {$Y_T$};
    \end{tikzpicture}%
    }
    \caption{\textbf{Memory-assisted sequential discrimination.} In round $j$ the referee draws an index $Z_j\sim\{p_i\}$ and sends the state $\sigma_{Z_j}$. The player applies a free operation $\Lambda$ jointly to $\sigma_{Z_j}$ and the current memory, outputs a guess $Y_j$, and forwards the updated memory (initialized in $\tau$) to the next round. No measurement acts across different rounds. The optimal asymptotic success rate is $R_{\mathrm{aux}}(\mathcal{S},\tau)$.}
    \label{fig:protocol}
\end{figure}

To begin, let us recall the setting and the sequential, memory-assisted discrimination game of Ref.~\cite{philip2025robustnessquantumdatahiding}, sketched in Fig.~\ref{fig:protocol}. The referee fixes the ensemble $\mathcal{S}=\{p_i,\sigma_i\}_{i=1}^N$ and, in each round, draws an index according to $\{p_i\}$ and sends the player the corresponding state. The player holds a quantum memory: a finite-dimensional auxiliary system initialized in some state $\tau$. In every round the player may apply an arbitrary free operation jointly to the freshly received state and the current memory, must output a guess for that round's index, and forwards the (updated) memory to the next round; crucially, no measurement may act jointly across distinct rounds. Writing $S_n$ for the number of correct guesses after $n$ rounds, a rate $r \in [0,1]$ is \emph{achievable} if, for every $\varepsilon>0$ and every $m>0$, there exists an $n \ge m$ such that $\Pr(S_n \ge r n) \ge 1-\varepsilon$, and the optimal assisted rate $R_{\mathrm{aux}}(\mathcal{S}, \tau)$ is the supremum of all achievable rates; by construction it depends on the initial memory $\tau$. This is precisely the configuration in which finite-dimensional catalysts and quantum memories are modeled~\cite{philip2025robustnessquantumdatahiding}: the memory may build up correlations with past rounds, but it cannot be externally replenished with fresh resource.

\begin{theorem}[Universal Limit of Finite-Dimensional Assistance]\label{thm:nogo_assistance}
Let $\mathcal{F}$ be a convex set of free states with interior and closed under tensor products, and let $\mathbb{M}$ be the measurements implementable by free
operations with free ancillas. We further require that free operations generate no resource even conditionally: applying a free operation to a free state and conditioning on any classical outcome it may produce again yields a free state. For any free-state ensemble $\mathcal{S}=\{p_i, \sigma_i\}_{i=1}^N$ with $\sigma_i \in \mathcal{F}$ and any finite-dimensional auxiliary state $\tau$,
\begin{equation}
    R_{\mathrm{aux}}(\mathcal{S}, \tau) = P_{\mathbb{M}}(\mathcal{S}).
\end{equation}
\end{theorem}

For a detailed proof of Theorem~\ref{thm:nogo_assistance}, see Supplementary Material. We provide a brief outline of proof here. The argument is by contradiction against the Helstrom bound. Suppose a finite-dimensional non-free state $\tau$ asymptotically improved the discrimination of $\mathcal{S}$. Pick a free state $\gamma \in \mathcal{F}$ that is non-orthogonal to $\tau$ such that $\|\tau-\gamma\|_1 < 2$.

Now consider the scenario where the initial state of the auxiliary quantum system is either $\tau$ or $\gamma$ with equal probability. Since $\mathcal{S}$ consists entirely of free states, the player can prepare it round after round using only free operations, and blindly runs the protocol optimized for $\tau$ across $n$ rounds. If the initial state of the auxiliary system is $\tau$, the empirical success rate converges to the advantaged value; if it is $\gamma$, the entire procedure is free and the rate cannot exceed $P_{\mathbb{M}}(\mathcal{S})$. Monitoring the empirical success rate identifies the initial state of the auxiliary system with probability approaching $1$ as $n \to \infty$, contradicting the Helstrom bound~\cite{helstrom1969quantum}. These arguments cover catalysis as well as quantum memory, provided the auxiliary system is finite-dimensional.

The addition of finite-dimensional auxiliary system can be thought of as persistent quantum side-information. Hence, the Theorem~\ref{thm:nogo_assistance} proves that generalized quantum data hiding schemes are robust against persistent quantum side-information. One can then ask the following question: Is this robustness unique to free state ensembles?

\paragraph{Memory Attacks on Non-Free-State Ensemble Discrimination}

To answer this question, we first make two assumptions about the underlying resource theory. The assumptions are as follows:
\begin{enumerate}
    \item Golden state: There exists a state, $\Phi$, such that augmenting $\mathcal{S}$ with $\Phi$ saturates the limit: $P_{\mathbb{M}}(\mathcal{S} \otimes \Phi) = P_{\mathrm{All}}(\mathcal{S})$, where $\mathcal{S} \otimes \Phi := \{p_i, \sigma_i \otimes \Phi\}_{i=1}^N$.

    \item Probabilistic distillation: For any $\epsilon,\tilde\epsilon>0$, there exist a resource state $\omega\notin\mathcal{F}$, a free state $\gamma\in\mathcal{F}$, an integer $k$, and a probabilistic free operation $\Lambda$ such that
    \begin{equation*}
        \Lambda(\omega^{\otimes k})=P_k\,\Phi^{\otimes k};\,\,\|\omega - \gamma\|_1 \le \epsilon;\,\,  P_k\ge 1-\tilde\epsilon.
    \end{equation*}
\end{enumerate}

The first assumption can be understood as existence of a single ``golden'' state that allows for all possible measurements to be performed using free operations. The second assumption requires that many copies of a state containing some resource can be probabilistically distilled into copies of the aforementioned golden state. Both hold in the resource theory of entanglement~\cite{philip2025robustnessquantumdatahiding,Hayashi02}. 

\begin{theorem}[Quantum Memory Attacks on Non-Free Ensembles]\label{thm4}
Consider a resource theory whose free measurements $\mathbb{M}$ are exactly those implementable by free operations with free ancillas. If it satisfies Assumptions 1 and 2, there exists a non-free state ensemble $\mathcal{S}$ and a finite-dimensional auxiliary state $\tau$ such that the memory-assisted asymptotic discrimination rate exceeds the optimal unassisted success probability:
$$R_{\mathrm{aux}}(\mathcal{S}, \tau) > P_{\mathbb{M}}(\mathcal{S}).$$
\end{theorem}

For a detailed proof of Theorem~\ref{thm4}, see Supplementary Material. We explain the key elements of the proof. Using the two assumptions, we construct a new ensemble of states by appending a resource state to a free state ensemble. We then show that the maximal success probability of distinguishing this constructed ensemble using free measurements is close to that of the free state ensemble. The following lemma is a key ingredient.
\begin{lemma}\label{lem:continuity}
    Let $\mathcal{S}=\{p_i, \sigma_i\}_{i=1}^N$ and $\mathcal{S}'=\{p_i, \sigma'_i\}_{i=1}^N$ be two state ensembles of the same dimension. If $\|\sigma_i - \sigma'_i\|_1 \le \epsilon$ for all $i$, then for any set of measurements $\mathbb{M}$,
    \begin{equation}
        |P_{\mathbb{M}}(\mathcal{S}) - P_{\mathbb{M}}(\mathcal{S}')| \le \epsilon.
    \end{equation}
\end{lemma}
The proof of this lemma is included in the Supplementary Material. We note that the above lemma holds for \emph{any} set of measurements, not just for restricted sets. Also, we would also like to point out that trace distance does not capture a state's resource potential across dimensions: a high-dimensional state arbitrarily close to $\mathcal{F}$ can nonetheless carry far more \emph{distillable} resource than a low-dimensional state far from it~\cite{philip2025robustnessquantumdatahiding,Hayashi02}. 

We then use this newly constructed ensemble to demonstrate a protocol that uses quantum memory to achieve higher distinguishability. The protocol for distinguishing the states in the new ensemble is as follows. The protocol divides the task into blocks of $k$ rounds. Starting with memory $\Phi^{\otimes k}$, the player consumes one $\Phi$ per round for an optimal guess while collecting the appended copies of $\omega$. At the end of each block, $\omega^{\otimes k}$ is distilled into a fresh $\Phi^{\otimes k}$: success enables optimal discrimination in the next block, while failure drops performance to $P_{\mathbb{M}}$ until another distillation succeeds. As the number of blocks grows, the overall rate is lower-bounded by $P_{\mathrm{All}}(\mathcal{S})(1-\tilde{\epsilon})$, where $P_{\mathrm{All}}(\mathcal{S})$ is the maximal success probability of distinguishing the ensemble using all possible measurements.

\paragraph{Discussion} In this work, we identify a sufficient condition on resource theories for the existence of \emph{free state ensembles} that exhibit the free-state discrimination gap. Theorem~\ref{thm1} shows that whenever the free measurements are confined to a proper closed convex subset of all measurements and the free states form a convex body with an interior, there is an ensemble of \emph{free} states that the free measurements cannot optimally distinguish. This recasts nonlocality without entanglement and nonstabilizerness without magic as two examples of a broader phenomenon in resource theories. It also indicates the existence of such free-state discrimination gaps  in essentially any resource theory meeting certain conditions, essentially providing a broad generalization of nonlocality without entanglement beyond the resource theory of entanglement. 

Theorems~\ref{thm:nogo_assistance} and~\ref{thm4} address some operational questions regarding free-state discrimination gap. Theorem~\ref{thm:nogo_assistance} shows that the free-state discrimination gap cannot be overcome with finite-dimensional catalyst or quantum memory assistance. This then implies that generalized quantum data hiding schemes constructed from free states are robust against attacks using persistent quantum side information. Theorem~\ref{thm4} shows that generalized quantum data hiding schemes that use \emph{resourceful} ensemble can be fully broken using quantum memory. It is thus the resource content of the data hiding ensemble that determines whether a quantum data hiding scheme remains robust. This shows a clear dichotomy between free and resourceful generalized quantum data hiding schemes. 

Our results also open up several open questions. The separating-hyperplane construction in Theorem~\ref{thm1} is fully constructive, suggesting a systematic SDP-based construction~\cite{Skrzypczyk_2023} of ensembles exhibiting a free-state discrimination gap in arbitrary resource theories.  The geometric proof of Theorem~\ref{thm1} relies on the free-state set being convex and having an interior. Whether an operational gap persists for resource theories with non-convex~\cite{Albarelli_2018} or not having an interior~\cite{Streltsov_2017} free sets remains unexplored. Also, since the argument in the proof of Theorem~\ref{thm1} does not rely on the structure of the Hilbert space, it can also be extended to generalized probabilistic theories~\cite{PLAVALA20231}.
We would like to point out that the set of free measurements need not match the chosen set of free states, for example: one could, for instance, restrict to stabilizer operations yet construct a gap-witnessing ensemble entirely from separable states. This could lead to interesting consequences for quantum computing.  
Finally, Theorem~\ref{thm4} shows that memory assistance can close the measurement gap for some non-free ensembles under two assumptions on the underlying resource theory. These are natural in entanglement theory but unclear elsewhere, particularly in the resource theory of magic. Identifying the precise boundary between memory-accessible ensembles and those that remain robustly hidden is a key direction for future work.

\textit{Acknowledgments}--- This work was supported by the National Science Centre Poland (Grant No. 2022/46/E/ST2/00115 and 2024/55/B/ST2/01590).

\bibliography{refs}

\newpage
\section{Supplementary Materials}

\subsection{Proof of Theorem~\ref{thm1}}
\label{app:universal_gap}

Fix any resourceful POVM $\mathcal{M}^*=(M_1^*,\dots,M_N^*)\notin\mathbb{M}$, and view
$N$-outcome POVMs as vectors in the real inner-product space
$V := \mathrm{Herm}(\mathcal{H})^{\oplus N}$ with
$\langle\mathcal{A},\mathcal{M}\rangle := \sum_{i=1}^N\tr[A_iM_i]$,
$\mathcal{A}=(A_1,\dots,A_N)$. Since $\mathbb{M}$ is closed and convex, the point
$\mathcal{M}^*\notin\mathbb{M}$ is strictly separated from it, and the strict
separation theorem~\cite{boyd2004convex} yields Hermitian operators
$A_1,\dots,A_N$ with
\begin{equation}\label{eqn:sep}
    \sum_{i=1}^N\tr[A_iM_i^*] \;>\; \sup_{\mathcal{M}\in\mathbb{M}}\sum_{i=1}^N\tr[A_iM_i].
\end{equation}
Geometrically, the tuple $\mathcal{A}=(A_1,\dots,A_N)$ is the normal of a hyperplane that strictly separates $\mathcal{M}^*$ from $\mathbb{M}$, as illustrated in Fig.~\ref{fig:sep}.

We turn $\mathcal{A}$ into a free-state ensemble by shifting the $A_i$ into the cone of free states (illustrated in Fig.~\ref{fig:geometry}). Since $\mathcal{F}$ has nonempty interior in the state space $\mathcal{D}(\mathcal{H})$, it is full-dimensional therein, so $\dim\mathcal{F}=d^{2}-1$ with $d:=\dim\mathcal{H}$. It follows that $\dim\operatorname{cone}(\mathcal{F})=d^{2}$, and hence $\operatorname{cone}(\mathcal{F})$ has nonempty interior in $\operatorname{Herm}(\mathcal{H})$. Pick an interior point $\gamma$ of $\mathrm{cone}(\mathcal{F})$ and a radius $\varepsilon>0$ small enough that the open ball $B(\gamma,\varepsilon)$ around $\gamma$ still lies inside the cone, $B(\gamma,\varepsilon)\subset\mathrm{cone}(\mathcal{F})$. For $t > \max_i\|A_i\|/\varepsilon$ we have $\gamma+A_i/t\in B(\gamma,\varepsilon)$, so by conic invariance
\begin{equation*}
    B_i := A_i + t\gamma = t\,(\gamma + A_i/t)\in\mathrm{int}(\mathrm{cone}(\mathcal{F})),
\end{equation*}
and in particular $B_i>0$. Set $\lambda_i := \tr[B_i] > 0$,
$\sigma_i := B_i/\lambda_i\in\mathcal{F}$, $Q := \sum_i\lambda_i$, and
$p_i := \lambda_i/Q$, giving a free-state ensemble
$\mathcal{S}=\{p_i,\sigma_i\}_{i=1}^N$. Because
$\sum_i\tr[t\gamma\,M_i] = t\tr[\gamma]$ is constant on POVMs, \eqref{eqn:sep} is
preserved under $A_i\mapsto B_i$, and the given measurement $\mathcal{M}^*$ obeys
\begin{align}
    P(\mathcal{S},\mathcal{M}^*)
    &= \frac{1}{Q}\sum_{i=1}^N\tr[B_iM_i^*]\nonumber\\
    &= \frac{1}{Q}\Big(\textstyle\sum_i\tr[A_iM_i^*] + t\tr[\gamma]\Big)\nonumber\\
    &> \frac{1}{Q}\Big(\sup_{\mathcal{M}\in\mathbb{M}}\textstyle\sum_i\tr[A_iM_i]
       + t\tr[\gamma]\Big)\nonumber\\
    &= P_{\mathbb{M}}(\mathcal{S}),
\end{align}
which completes the proof.
\begin{figure}[t]
    \centering
    \resizebox{\columnwidth}{!}{%
    \begin{tikzpicture}[>=Stealth, font=\sffamily]

    \draw[thick, fill=gray!8, draw=black!60, rounded corners=22pt]
        (-1.3,-0.6) rectangle (10.5,5.6);
    \node[anchor=north west, text=black!70] at (-1.05,5.45)
        {\large $\mathrm{POVM}_N(\mathcal{H})$};

    \draw[thick, fill=gray!30, draw=black, rounded corners=10pt]
        (0.5, 1.5) -- (1.0, 3.5) -- (3.0, 4.0) -- (4.5, 2.5) -- (3.0, 1.0) -- cycle;
    \node[text=black] at (2.4, 2.5) {\Large $\mathbb{M}$};

    \coordinate (Mstar) at (5.8, 4.4);
    \fill[black] (Mstar) circle (2.5pt) node[above left, yshift=1pt] {\Large $\mathcal{M}^*$};

    \draw[very thick, dashed, black] (3.1, 5.3) -- (6.9, 0.25);

    \draw[->, very thick, black] (5.74, 1.79) -- (6.59, 2.43);
    \draw[thick, black] (5.52, 2.08) -- (5.84, 2.32) -- (6.08, 2.00);
    \node[right, xshift=3pt, yshift=-2pt] at (6.59, 2.43)
        {\large $\mathcal{A} = (A_1, \dots, A_N)$};
    \end{tikzpicture}%
    }
    \caption{\textbf{Separating a resourceful measurement from the free set.}
The free measurements $\mathbb{M}$ form a convex region that occupies only part
of the full set of measurements $\mathrm{POVM}_N(\mathcal{H})$, so any resourceful
measurement $\mathcal{M}^*\notin\mathbb{M}$ lies outside it and can be separated
from $\mathbb{M}$ by a hyperplane (dashed). Its normal $\mathcal{A}=(A_1,\dots,A_N)$ plays the role of a witness: after a suitable shift, these operators become a free-state ensemble that $\mathcal{M}^*$ distinguishes better than any free measurement can.}
    \label{fig:sep}
\end{figure}
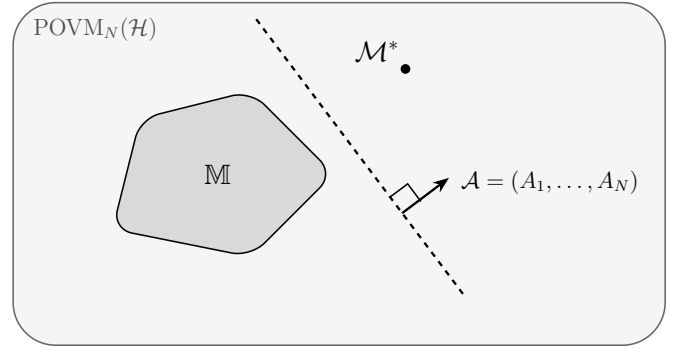
\begin{figure}[t]
    \centering
     \begin{tikzpicture}[>=Stealth, scale=1, font=\sffamily]

    \coordinate (Apex) at (0, 4.5);

    \coordinate (BallCenter) at (0, 2.2);

    \coordinate (A1) at (3.0, 4.2);
    \coordinate (A2) at (-1.0, 4.8);
    \coordinate (AN) at (-3.5, 3.2);

    \coordinate (B1) at ($ (A1)!0.88!(BallCenter) $);
    \coordinate (B2) at ($ (A2)!0.86!(BallCenter) $);
    \coordinate (BN) at ($ (AN)!0.90!(BallCenter) $);

    \coordinate (S1) at (intersection of Apex--B1 and -4,-0.2--4,-0.2);
    \coordinate (S2) at (intersection of Apex--B2 and -4,-0.1--4,-0.1);
    \coordinate (SN) at (intersection of Apex--BN and -4,-0.3--4,-0.3);

    \draw[thick] (-2.8, 0) arc (180:360:2.8cm and 0.8cm); 
    \draw[thick, dashed, opacity=0.6] (2.8, 0) arc (0:180:2.8cm and 0.8cm);
    \node at (0.2, -0.4) {\Large $\mathcal{F}$};

    \draw[thick] (-2.8, 0) -- (Apex);
    \draw[thick] (2.8, 0) -- (Apex);

    \draw[thick] (BallCenter) circle (0.75cm);

    \draw[dashed, thick] (A1) -- (BallCenter);
    \draw[dashed, thick] (A2) -- (BallCenter);
    \draw[dashed, thick] (AN) -- (BallCenter);

    \draw[dashed, opacity=0.6] (Apex) -- (B1);
    \draw[dashed, opacity=0.6] (Apex) -- (B2);
    \draw[dashed, opacity=0.6] (Apex) -- (BN);
    \draw[dashed, thick] (B1) -- (S1);
    \draw[dashed, thick] (B2) -- (S2);
    \draw[dashed, thick] (BN) -- (SN);

    \path (A2) -- (AN) node[midway, sloped] {\Large $\dots$};
    
    \path (B2) -- (BN) node[midway, sloped] {$\dots$};
    
    \path (S2) -- (SN) node[midway, sloped] {$\dots$};

    \fill (A1) circle (1.5pt) node[right, xshift=2pt] {\large $A_1$};
    \fill (A2) circle (1.5pt) node[above left, yshift=2pt] {\large $A_2$};
    \fill (AN) circle (1.5pt) node[above left, yshift=2pt] {\large $A_N$};

    \fill (BallCenter) circle (1.5pt) node[right, xshift=2pt, yshift=-2pt] {\large $\gamma$};

    \fill (B1) circle (1.5pt);
    \fill (B2) circle (1.5pt);
    \fill (BN) circle (1.5pt);

    \fill (S1) circle (1.5pt) node[below right, xshift=-2pt, yshift=2pt] {$\sigma_1$};
    \fill (S2) circle (1.5pt) node[below left, xshift=2pt, yshift=2pt] {$\sigma_2$};
    \fill (SN) circle (1.5pt) node[below, yshift=-2pt] {$\sigma_N$};

    \end{tikzpicture}

    \caption{\textbf{Construction of the free-state witnessing ensemble (proof of Theorem~\ref{thm1}).} The separating normals $A_i$ may fall outside the cone of free states $\mathrm{cone}(\mathcal{F})$. Shifting each by a large multiple of an interior anchor $\gamma$, $B_i = A_i + t\gamma$, lands it in a ball inside the cone; projecting $B_i$ from the cone's apex onto the unit-trace cross-section then gives the free state $\sigma_i = B_i/\tr[B_i]$, with priors $p_i \propto \tr[B_i]$.}
    \label{fig:geometry}
\end{figure}
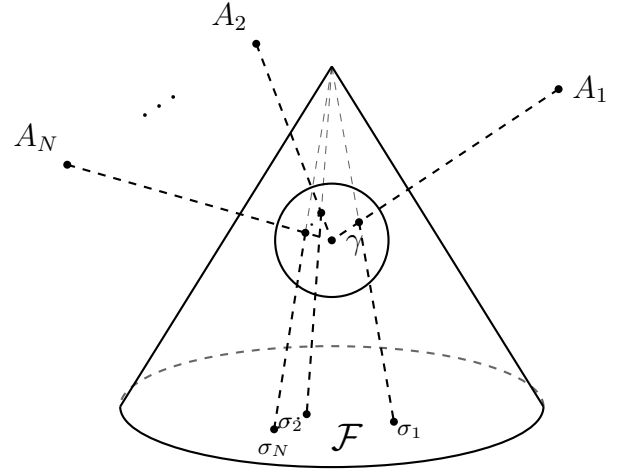

\subsection{Proof of Theorem~\ref{thm:nogo_assistance}}
\label{app:nogo}
Throughout, we use the hypotheses of Theorem~\ref{thm:nogo_assistance}, stated with it in the main text: the free-state set $\mathcal{F}$ is closed under tensor products; free operations generate no resource even conditionally, i.e., applying a free operation to a free state and conditioning on any classical outcome it may produce again yields a free state; and the free measurements $\mathbb{M}$ are exactly those implementable by free operations with free ancillas.

Our proof proceeds by contradiction. Assume that there exists a finite-dimensional auxiliary state $\tau$ that provides an asymptotic advantage in distinguishing the free-state ensemble $\mathcal{S} = \{p_i, \sigma_i\}_{i=1}^N$, such that $R_{\mathrm{aux}}(\mathcal{S}, \tau) > P_{\mathbb{M}}(\mathcal{S})$. 

Clearly, $\tau \notin \mathcal{F}$, otherwise the joint operations would remain entirely free and the success probability would be trivially bounded by $P_{\mathbb{M}}(\mathcal{S})$. Since the set of free states $\mathcal{F}$ is a convex body with interior, it contains a full-rank state. This ensures that we can always find a free state $\gamma \in \mathcal{F}$ that is non-orthogonal to $\tau$ (i.e., $\|\tau - \gamma\|_1 < 2$). Let us set up a discrimination task where the referee prepares an initial auxiliary state $\eta_1 \in \{\tau, \gamma\}$ with uniform prior probabilities $p_\tau = p_\gamma = 1/2$.

Choose a strictly positive constant $\delta$ such that $0 < \delta < R_{\mathrm{aux}}(\mathcal{S}, \tau) - P_{\mathbb{M}}(\mathcal{S})$. We define the following sequential protocol:
\begin{enumerate}
    \item[\textbf{0)}] Initialize round counter $j=1$. 
    \item[\textbf{1)}] The referee selects $Z_j \in \{1, \dots, N\}$ according to the distribution $\{p_i\}$ and prepares $\sigma_{Z_j} \in \mathcal{F}$. 
    \item[\textbf{2)}] The player applies the optimal sequence of free operations on the joint system $\sigma_{Z_j} \otimes \eta_j$ to obtain a guess $Y_j$.
    \item[\textbf{3)}] Define the success indicator $X_j = 1$ if $Y_j = Z_j$, and $X_j = 0$ otherwise.
    \item[\textbf{4)}] Update the auxiliary state to $\eta_{j+1}$ via the free operation, set $j \rightarrow j+1$, and repeat.
\end{enumerate}

After $n$ rounds, the empirical number of successes is $S_n = \sum_{j=1}^n X_j$. The player uses the following rule to identify $\eta_1$: guess $\tau$ if $\frac{S_n}{n} - P_{\mathbb{M}}(\mathcal{S}) \ge \delta$; otherwise, guess $\gamma$.

\textit{Case 1: $\eta_1 = \tau \notin \mathcal{F}$.---}
Since $R_{\mathrm{aux}}(\mathcal{S}, \tau) > P_{\mathbb{M}}(\mathcal{S}) + \delta$, we fix an achievable rate $r$ with $P_{\mathbb{M}}(\mathcal{S}) + \delta < r \le R_{\mathrm{aux}}(\mathcal{S}, \tau)$. By the definition of an achievable rate, for every $\varepsilon > 0$ and every $m > 0$ there exists an $n \ge m$ such that $\Pr(S_n \ge r n) \ge 1 - \varepsilon$. Since $r > P_{\mathbb{M}}(\mathcal{S}) + \delta$, the event $\{S_n \ge rn\}$ is contained in $\{S_n/n - P_{\mathbb{M}}(\mathcal{S}) \ge \delta\}$; hence for every such $n$, the probability of correctly identifying $\tau$ is bounded by:
\begin{equation}
    P_{\mathrm{corr}}(\tau) = \Pr\left( \frac{S_n}{n} - P_{\mathbb{M}}(\mathcal{S}) \ge \delta \right) \ge 1 - \varepsilon.
\end{equation}

\textit{Case 2: $\eta_1 = \gamma \in \mathcal{F}$.---}
Since $\sigma_{Z_j} \in \mathcal{F}$, $\gamma \in \mathcal{F}$, and all applied operations are free, the updated auxiliary state $\eta_j$ remains within the free states for all $j$. Consequently, the effective measurement acting solely on the system at any step belongs to the restricted set $\mathbb{M}$. By the definition of $P_{\mathbb{M}}(\mathcal{S})$, the conditional probability of success is bounded:
\begin{equation}
    \Pr(X_{j+1}=1 \mid X_1, \dots, X_j) \le P_{\mathbb{M}}(\mathcal{S}).
\end{equation}
Let us define the sequence of random variables $C_j := S_j - j P_{\mathbb{M}}(\mathcal{S})$. It is straightforward to verify that:
\begin{align}
    &\mathbb{E}[C_{j+1} \mid C_j, \dots, C_1] \nonumber \\
    &= C_j - P_{\mathbb{M}}(\mathcal{S}) + \mathbb{E}[X_{j+1} \mid X_1, \dots, X_j] \nonumber \\
    &\le C_j.
\end{align}
With $|C_{j} - C_{j-1}| \le 1$, the sequence $\{C_j\}$ forms a supermartingale. Applying Azuma's inequality \cite{azuma1967weighted} yields:
\begin{equation}
    \Pr\left( \frac{S_n}{n} - P_{\mathbb{M}}(\mathcal{S}) \ge \delta \right) = \Pr(C_n \ge n\delta) \le \exp\left(-\frac{n\delta^2}{2}\right).
\end{equation}
Thus, the probability of correctly identifying $\gamma$ is:
\begin{equation}
    P_{\mathrm{corr}}(\gamma) \ge 1 - \exp\left(-\frac{n\delta^2}{2}\right).
\end{equation}

The overall probability of correctly discriminating between $\tau$ and $\gamma$ using this $n$-round protocol is:
\begin{equation}\label{eqn:theorem_2_p_succ}
    P_{\mathrm{corr}} \ge \frac{1}{2} (1 - \varepsilon) + \frac{1}{2} \left[ 1 - \exp\left(-\frac{n\delta^2}{2}\right) \right].
\end{equation}
However, according to the Helstrom bound, the fundamental limit for discriminating single copies of the non-orthogonal states $\tau$ and $\gamma$ is strictly bounded away from 1:
\begin{equation}
    P_{\mathrm{opt}}(\tau, \gamma) = \frac{1}{2} + \frac{1}{4} \|\tau - \gamma\|_1 < 1.
\end{equation}
Since, $\varepsilon$ and $n$ can be varied arbitrarily, pick $\varepsilon$ such that $0 < \varepsilon < 1 - P_{\mathrm{opt}}(\tau, \gamma)$ and choose $n$ large enough such that $\exp(-n\delta^2/2) \le \varepsilon$. Applying these to \eqref{eqn:theorem_2_p_succ}, we get $P_{\mathrm{corr}} \ge 1 - \varepsilon > P_{\mathrm{opt}}(\tau, \gamma)$. This contradicts the Helstrom bound, so our initial assumption is false and $R_{\mathrm{aux}}(\mathcal{S}, \tau) \le P_{\mathbb{M}}(\mathcal{S})$. The reverse inequality is immediate, since discarding $\tau$ and applying a near-optimal free measurement in each round achieves every rate below $P_{\mathbb{M}}(\mathcal{S})$; hence $R_{\mathrm{aux}}(\mathcal{S}, \tau) = P_{\mathbb{M}}(\mathcal{S})$.

\subsection{Proof of Lemma~\ref{lem:continuity}}
\label{app:memory}

By the definition of the supremum $P_{\mathbb{M}}(\mathcal{S})$, for any arbitrarily small $\eta > 0$, there exists a measurement $\{M_i^\eta\}_{i=1}^N \in \mathbb{M}$ whose success probability approaches the optimal value within $\eta$:
\begin{equation}
    \sum_{i=1}^N p_i \tr[M_i^\eta \sigma_i] \ge P_{\mathbb{M}}(\mathcal{S}) - \eta.
\end{equation}
We can analyze the performance of this specific measurement when applied to the counterpart ensemble $\mathcal{S}'$:
\begin{align}
    P_{\mathbb{M}}(\mathcal{S}) - \eta &\le \sum_{i=1}^N p_i \tr[M_i^\eta \sigma_i] \nonumber \\
    &= \sum_{i=1}^N p_i \tr[M_i^\eta \sigma'_i] + \sum_{i=1}^N p_i \tr[M_i^\eta (\sigma_i - \sigma'_i)] \nonumber \\
    &\le P_{\mathbb{M}}(\mathcal{S}') + \sum_{i=1}^N p_i |\tr[M_i^\eta (\sigma_i - \sigma'_i)]| \nonumber \\
    &\le P_{\mathbb{M}}(\mathcal{S}') + \sum_{i=1}^N p_i \|\sigma_i - \sigma'_i\|_1 .
\end{align}
In the first inequality, we used the fact that $\sum_i p_i \tr[M_i^\eta \sigma'_i]$ is the success probability of a particular valid measurement in $\mathbb{M}$ for $\mathcal{S}'$, which is inherently upper bounded by its supremum $P_{\mathbb{M}}(\mathcal{S}')$. In the last inequality, we used the trace-norm bound $|\tr[M^\eta_i \Delta]| \le \|M^\eta_i\|_\infty \|\Delta\|_1 \le \|\Delta\|_1$, which holds because the operator norm of any POVM element satisfies $\|M^\eta_i\|_\infty \le 1$.

Since we are given that $\|\sigma_i - \sigma'_i\|_1 \le \epsilon$ for all $i$, we have:
\begin{equation}
    P_{\mathbb{M}}(\mathcal{S}) - P_{\mathbb{M}}(\mathcal{S}') \le \epsilon + \eta.
\end{equation}
Because $\eta > 0$ is strictly arbitrary, taking the limit $\eta \to 0^+$ yields:
\begin{equation}
    P_{\mathbb{M}}(\mathcal{S}) - P_{\mathbb{M}}(\mathcal{S}') \le \epsilon.
\end{equation}

By exchanging the roles of $\mathcal{S}$ and $\mathcal{S}'$, a completely symmetric argument gives $P_{\mathbb{M}}(\mathcal{S}') - P_{\mathbb{M}}(\mathcal{S}) \le \epsilon$. Combining these concludes the proof:
\begin{equation}
    |P_{\mathbb{M}}(\mathcal{S}) - P_{\mathbb{M}}(\mathcal{S}')| \le \epsilon.
\end{equation}

\subsection{Proof of Theorem~\ref{thm4}}

\textit{The construction---}
By Theorem~\ref{thm1} fix a free ensemble $\mathcal{S}=\{p_i,\sigma_i\}_{i=1}^N$ with a strict gap $P_{\mathbb{M}}(\mathcal{S})<p^*:=P_{\mathrm{All}}(\mathcal{S})$. Let $\Phi$, $\omega\notin\mathcal{F}$, $\gamma\in\mathcal{F}$, $k$, and $\Lambda$ be such that Assumptions~1 \& 2 are satisfied. The players try to distinguish $\mathcal{S}\otimes\omega=\{p_i,\sigma_i\otimes\omega\}_{i=1}^N$ using the fixed memory $\tau=\Phi^{\otimes k}$, in blocks of $k$ rounds. In each round, they spend one $\Phi$ to distinguish the received $\sigma_{Z_j}$ with probability  $p^*$ and store the appended $\omega$. After $k$ rounds, they distill the collected $\omega^{\otimes k}$ back to $\Phi^{\otimes k}$ via $\Lambda$ with a significantly high probability, refueling for the next block. Since $k$ is fixed, $\tau$ is finite-dimensional.

\textit{Single-shot gap is preserved---}
By Assumption 2, for any $\epsilon > 0$, there exists a resource state $\omega$ and a free state $\gamma \in \mathcal{F}$ such that $\|\omega - \gamma\|_1 \le \epsilon$. Because $\gamma$ is a free state, it provides no advantage under restricted operations: $P_{\mathbb{M}}(\mathcal{S} \otimes \gamma) = P_{\mathbb{M}}(\mathcal{S})$. Applying the continuity bound from Lemma~\ref{lem:continuity}, we immediately obtain:
\begin{equation*}
    |P_{\mathbb{M}}(\mathcal{S} \otimes \omega) - P_{\mathbb{M}}(\mathcal{S})| \le \|\omega - \gamma\|_1 \le \epsilon.
\end{equation*}
\textit{Memory collapses the gap---}
Choose the block length $k$ as in Assumption~2 and perform $n=kt$ rounds, viewed as $t$ independent blocks of $k$ rounds each, as in Ref.~\cite{philip2025robustnessquantumdatahiding}. Within a block the player starts from $\Phi^{\otimes k}$, spends one $\Phi$ per round to attain the optimal success probability $p^*$ on the received state (Assumption~1), and at the end of the block distils the collected $\omega^{\otimes k}$ into a fresh $\Phi^{\otimes k}$ for the next block; by Assumption~2 this distillation succeeds with probability $P_k\ge1-\tilde\epsilon$. Writing $Y_j\in[0,1]$ for the success fraction of block $j$, so that $S_n=k\sum_{j=1}^{t}Y_j$, the blocks are independent and each attains
\begin{equation*}
    \mathbb{E}[Y_j]\ge p^*(1-\tilde\epsilon),
\end{equation*}
the factor $1-\tilde\epsilon$ accounting for the blocks in which the distillation fails, whose contribution we lower-bound by zero. Define $r=p^*(1-\tilde\epsilon)-\delta$ for some $\delta>0$. Since the $\{Y_j\}$ are independent and take values in $[0,1]$, Hoeffding's inequality~\cite{hoeffding1963probability} shows that for any $\varepsilon,m>0$ we can choose a $t$ with $kt\ge m$ large enough that
\begin{equation*}
    \Pr(S_{kt}\ge r\,kt)\ge 1-\varepsilon.
\end{equation*}
As $\delta>0$ is arbitrary and $P_{\mathrm{All}}(\mathcal{S}\otimes\omega)=p^*$, every rate below $p^*(1-\tilde\epsilon)$ is achievable, so
\begin{equation*}
    R_{\mathrm{aux}}(\mathcal{S}\otimes\omega,\,\Phi^{\otimes k})\ge p^*(1-\tilde\epsilon)=P_{\mathrm{All}}(\mathcal{S}\otimes\omega)(1-\tilde\epsilon).
\end{equation*}
 
Choosing $\epsilon,\tilde{\epsilon}$ small enough that $\epsilon+\tilde{\epsilon}\,p^*<p^*-P_{\mathbb{M}}(\mathcal{S})$, the single-shot bound and the rate bound combine to give
\begin{equation*}
    R_{\mathrm{aux}}(\mathcal{S}\otimes\omega,\tau) \ge p^*(1-\tilde{\epsilon}) > P_{\mathbb{M}}(\mathcal{S}) + \epsilon \ge P_{\mathbb{M}}(\mathcal{S}\otimes\omega),
\end{equation*}
completing the proof.
\end{document}